\documentclass[aps,prab,twocolumn,
superscriptaddress,
floatfix,
letter,
longbibliography,
10pt]{revtex4-2}

\usepackage{physics}
\usepackage{standalone}
\usepackage{graphicx}
\usepackage[utf8]{inputenc}
\usepackage{amssymb}
\usepackage{xcolor}
\usepackage{dirtytalk}
\usepackage[margin=1in]{geometry}
\usepackage{hyperref}
\usepackage{cleveref}
\usepackage{comment}
\usepackage{float}

\newcommand{\up}{\uparrow}
\newcommand{\down}{\downarrow}

\begin{document}

\title{Comparing Symmetrized Determinant Neural Quantum States for the Hubbard Model}

\author{Louis Sharma}
\email{These two authors contributed equally.}
\affiliation{CPHT, CNRS, Ecole Polytechnique, Institut Polytechnique de Paris, 91120 Palaiseau, France}
\affiliation{Collège de France, Université PSL, 11 place Marcelin Berthelot, 75005 Paris, France}

\author{Ahmedeo Shokry}
\email{These two authors contributed equally.}
\affiliation{CPHT, CNRS, Ecole Polytechnique, Institut Polytechnique de Paris, 91120 Palaiseau, France}
\affiliation{Collège de France, Université PSL, 11 place Marcelin Berthelot, 75005 Paris, France}
\affiliation{Inria Paris-Saclay, Bâtiment Alan Turing, 1, rue Honoré d’Estienne d’Orves – 91120 Palaiseau}
\affiliation{LIX, CNRS, École polytechnique, Institut Polytechnique de Paris, 91120 Palaiseau, France}

\author{Rajah Nutakki}
\affiliation{CPHT, CNRS, Ecole Polytechnique, Institut Polytechnique de Paris, 91120 Palaiseau, France}
\affiliation{Collège de France, Université PSL, 11 place Marcelin Berthelot, 75005 Paris, France}
\affiliation{Inria Paris-Saclay, Bâtiment Alan Turing, 1, rue Honoré d’Estienne d’Orves – 91120 Palaiseau}
\affiliation{LIX, CNRS, École polytechnique, Institut Polytechnique de Paris, 91120 Palaiseau, France}

\author{Olivier Simard}
\affiliation{CPHT, CNRS, Ecole Polytechnique, Institut Polytechnique de Paris, 91120 Palaiseau, France}
\affiliation{Collège de France, Université PSL, 11 place Marcelin Berthelot, 75005 Paris, France}

\author{Michel Ferrero}
\affiliation{CPHT, CNRS, Ecole Polytechnique, Institut Polytechnique de Paris, 91120 Palaiseau, France}
\affiliation{Collège de France, Université PSL, 11 place Marcelin Berthelot, 75005 Paris, France}

\author{Filippo Vicentini}
\affiliation{CPHT, CNRS, Ecole Polytechnique, Institut Polytechnique de Paris, 91120 Palaiseau, France}
\affiliation{Collège de France, Université PSL, 11 place Marcelin Berthelot, 75005 Paris, France}
\affiliation{Inria Paris-Saclay, Bâtiment Alan Turing, 1, rue Honoré d’Estienne d’Orves – 91120 Palaiseau}
\affiliation{LIX, CNRS, École polytechnique, Institut Polytechnique de Paris, 91120 Palaiseau, France}

\begin{abstract}
Accurate simulations of the Hubbard model are crucial to understanding strongly correlated phenomena, where small energy differences between competing orders demand high numerical precision. In this work, Neural Quantum States are used to probe the strongly coupled and underdoped regime of the square-lattice Hubbard model.
We systematically compare the Hidden Fermion Determinant State and the Jastrow-Backflow ansatz, parametrized by a Vision Transformer, finding that in practice, their accuracy is similar.
We also test different symmetrization strategies, finding that output averaging yields the lowest energies, though it becomes costly for larger system sizes.
On cylindrical systems, we consistently observe filled stripes.
On the torus, our calculations display features consistent with a doped Mott insulator, including antiferromagnetic correlations and suppressed density fluctuations.
Our results demonstrate both the promise and current challenges of neural quantum states for correlated fermions.
\end{abstract}
\maketitle

\section{Introduction}
The Hubbard model~\cite{arovas2022, qin2022} occupies a central role in the understanding of correlated electrons in condensed matter. 
It is thought to host a broad range of phases of matter from antiferromagnetism, spin- and charge-density waves, to $d$-wave superconductivity and even topologically-ordered spin liquids.
Much interest stems from its ability to capture essential qualitative properties of experimental correlated quantum systems, such as the high-$T_{\text{c}}$ cuprate~\cite{BednorzMuller1986,Keimer2015} and nickelate~\cite{Kitatani_2020} superconductors or organic charge-transfer crystals $\kappa$-(BEDT-TTF)$_2$~\cite{PhysRevLett.91.107001,Yamashita2008,Yamashita2009}. 
Despite this, even on the square lattice, reaching a consensus on its phase diagram across the entire parameter space (hoppings, interaction strengths and fillings) remains a considerable challenge.
Many numerical methods are under active development to tackle this and similar problems, including DMFT~\cite{RevModPhys.68.13,RevModPhys.77.1027,RevModPhys.90.025003}, DMRG~\cite{PhysRevB.48.10345,schollwock2011,stoudenmire2012}, PEPS~\cite{kraus2010,cirac2021a}, quantum Monte Carlo (QMC)~\cite{PhysRevLett.81.2514, sugiyama1986, zhang2013, becca2017, kozik2010} and determinant expansions~\cite{giuliani2025chemistry} with recent notable successes in tackling challenging points of the phase diagram~\cite{leblanc2015solutions, qin2016, zheng2017stripe,simard2019, qin2020, xiao2023, xu2024, simkovic2024, liu2025accurate}.

At the same time, the use of neural networks as variational wavefunctions within a variational Monte Carlo framework (neural quantum states)~\cite{carleo2017solving,Vicentini2021}, has emerged as a promising method for studying ground state properties of quantum many-body systems and non-equilibrium dynamics~\cite{Gravina2025,Schmitt2022,sinibaldi2024time,parnes2025nuclear,Nys_2024,Nys2024Thermofield}.
In particular, for ground states of frustrated spin systems, neural quantum states (NQS) have achieved unprecedented accuracy~\cite{nomura2021, Chen2024,rende2024,viteritti2025,duric2024}, building consensus around the nature of ground states in these challenging models. 

The method has also been applied to lattice fermionic systems~\cite{nomura2017, romero2025, lange2024simulating, zhang2025}, with particular focus on the $t' = 0$ square-lattice Hubbard model in order to benchmark different approaches, such as the neural backflow~\cite{luo2019backflow}, hidden fermion determinant state~\cite{robledo2022fermionic}, hidden fermion Pfaffian state~\cite{chen2025neuralnetworkaugmentedpfaffianwavefunctions} or transformer wavefunction~\cite{gu2025solving}.
However, a fair comparison among these architectures does not exist.
An advantage of NQS over other methods is the ability to simulate lattices with periodic boundaries, potentially helping to alleviate the difficulties in extrapolating the results of simulations on finite-size systems to the thermodynamic limit.

The aim of this work is to take the lessons learnt from the successful application of NQS to spin systems (neural network architectures, optimization strategies, use of lattice symmetries) and investigate their applicability to fermionic systems.
Furthermore, whilst the similarity between the backflow and hidden fermion determinant approaches can be shown explicitly~\cite{liu2024unifying}, their relative strengths and weaknesses in practice, particularly in combination with lattice symmetries have not been systematically studied.
We perform such a study here.

The remainder of the paper is organized as follows.
In Sec.~\ref{sec:Hubbard_model}, we present the Hubbard model and in Sec.~\ref{sec:model_and_methods} the NQS ansätze employed to study its ground state properties. 
Then, in Sec.~\ref{sec:numerical_results}, we present and discuss results on $8 \times L$ lattices with open and periodic boundary conditions. 

\section{Hubbard model}
\label{sec:Hubbard_model}
We consider the single-band Fermi-Hubbard model on an $M$-site 2d square lattice with nearest neighbor hopping. 
The Hamiltonian is given by
\begin{equation}
    \mathcal{\hat H} = - t\sum_{\expval{ij}, \sigma}  (\hat c^\dag_{i\sigma} \hat c_{j\sigma} + h.c.) + U \sum_i \hat n_{i\up}\hat n_{i\down},
\end{equation}
where $t$ is the hopping matrix element between nearest neighbor sites $\expval{ij}$, $\sigma = \{\up, \down\}$ is the spin index, $\hat c^\dag_{i\sigma}$ and $\hat c_{i\sigma}$ are fermionic operators that add or remove electrons from the single particle states $\ket{i\sigma}.$ They satisfy the canonical anticommutation relations
\begin{equation}
    \acomm*{\hat c_{i\sigma}}{\hat c^\dag_{j\sigma'}} = \delta_{ij} \delta_{\sigma\sigma'}, \quad \acomm{\hat c_{i\sigma}}{\hat c_{j\sigma'}} = 0.
\end{equation}
Finally, $U$ is the on-site Coulomb repulsion and $\hat n_{i\sigma} = \hat c^\dag_{i\sigma} \hat c_{i\sigma}$ is the number operator. In this study, we work in sectors of the Hilbert space corresponding to a fixed particle number $N = N_\up+ N_\down$ and fixed magnetization $2S^z = N_\up - N_\down.$
The corresponding electron (hole) density is $n = N/M$ ($n_h = 1 - n$).

We focus on the Hubbard model at strong coupling ($U/t = 8$) and hole-doping $n_h = 1/8$, a particular challenge due to the severity of the fermionic sign problem~\cite{loh1990}, non-perturbative parameter regime and presence of competing ground states~\cite{qin2022}. 
At half-filling, the absence of a sign-problem means that QMC can provide numerically exact solutions, showing that the ground state is a N\'eel antiferromagnet~\cite{varney2009, qin2016}.
Upon doping, there is competition between a uniform d-wave superconducting state and stripe states.
In a stripe state holes cluster together along one direction, forming a line defect in the antiferromagnetic spin background and a periodic modulation in hole density with wavelength $\lambda$~\cite{vojta2009, simard2026chargespinorderstuvj}.
Several results~\cite{qin2016, ehlers2017, zheng2017stripe,ido2018, darmawan2018, huang2018,  jiang2019, tocchio2019, qin2022, liu2025accurate, gu2025solving} find a $\lambda = 8$ filled stripe (with periodicity $2\lambda$ in the spin correlations) as the ground state for $U/t = 8, n_h = 1/8$.

\section{Method}
\label{sec:model_and_methods}
\subsection{Determinant states for fermionic wavefunctions}

A generic second-quantized many-body wavefunction can be written as
\begin{equation}
    \ket{\psi} = \sum_{\vb*n} \psi(\vb*n) \ket{\vb*n}
\end{equation}
where $\ket{\vb*n}$ denotes states in the Fock basis.
A class of fermionic NQS developed in the past few years are those based on determinants, having the advantage of a straightforward connection to a mean-field Slater determinant and indications that this structure is advantageous for obtaining strongly-correlated ground states~\cite{Denis_2025}.

Whilst other types of fermionic NQS ans\"atze exist, such as those based on Pfaffians~\cite{PhysRevB.96.205152,chen2025neuralnetworkaugmentedpfaffianwavefunctions}, our focus is on the determinant states which we define in the following.\\

\noindent \textbf{Neural Network Backflow:} Originally introduced in Ref.~\cite{luo2019backflow}, the neural network backflow approach parameterizes an additive transformation of the single-particle orbitals, which we use to construct a linear combination of Slater determinants,
\begin{equation}
    \psi_{\text{Bf}}(\vb*n) =  \sum_k C_k \prod_{\sigma}\text{det}\bigg([\vb*\Phi^{\sigma, k}(\vb*n)]_{\vb*n}\bigg).
\end{equation}
where $\vb*\Phi^{\sigma,k}$ is an $M \times N_\sigma$ matrix and $[\vb\Phi^{\sigma}(\vb*n)]_{\vb*n}$ denotes the matrix obtained by selecting the $N_\sigma$ rows corresponding to the occupied orbitals in the configuration $\vb*n$, where $N_{\sigma}$ is the number of fermions with spin $\sigma$.
The coefficients $C_k$ are free parameters.
The matrix entries $\vb*\Phi^{\sigma}_{ij}$ are given by the sum of a ``mean-field" orbital and an input-dependent correction
\begin{equation}
    \vb*\Phi^{\sigma}_{ij}(\vb*n) = \vb\Phi_{ij}^{\sigma, (0)} + \tilde{\vb*\Phi}_{ij}^{\sigma} (\vb*n).
\end{equation}
Here, $\vb\Phi_{ij}^{\sigma, (0)}$ is the $i$-th variational mean-field orbital evaluated at the position of the $j$-th particle with spin $\sigma$, while $\tilde{\vb\Phi}_{ij}^{\sigma} (\vb*n)$ is the input-dependent correction obtained from a neural network.

We also include a Jastrow density-density correlation term.
\begin{equation}
    J(\vb*n) = \exp\left( -\frac{1}{2} \sum_{i,j}  n_i W_{ij} n_j \right),
\end{equation}
with $W$ an upper triangular matrix. The full variational wavefunction is then written as
\begin{equation}
    \psi_{\text{JBf}}(\vb*n) = J(\vb*n) \cdot \psi_{\text{Bf}}(\vb*n),
\end{equation}
which we call the Jastrow-Backflow (JBf).
The set of variational parameters includes those of the neural network backflow, the mean-field orbitals, the Jastrow matrix $W$, and the Slater coefficients $C_k$. \\

\begin{figure*}[th!]
    \centering
    \includegraphics[width=1.\linewidth]{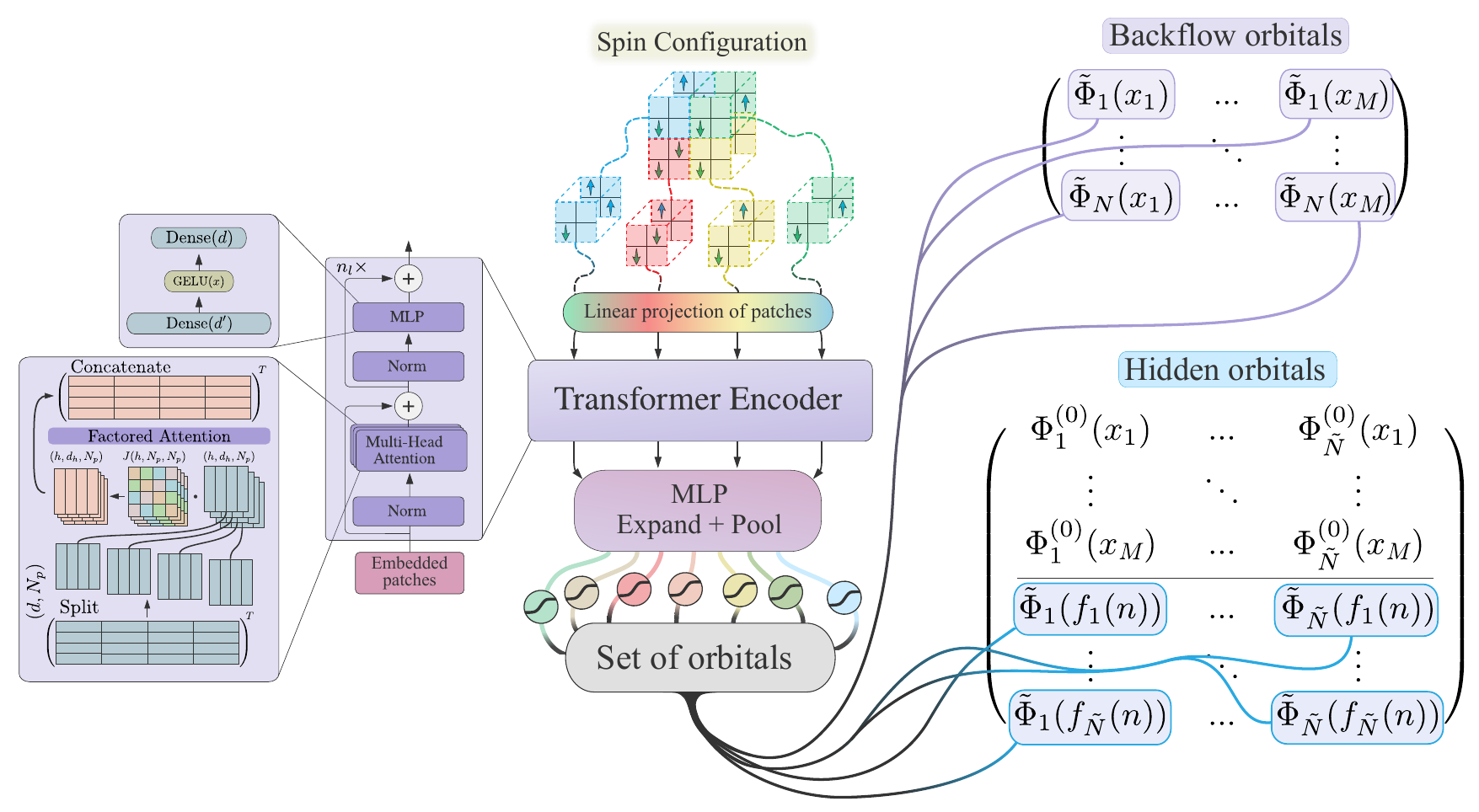}
    \caption{Schematic representation of the fermionic neural network wavefunctions discussed in this manuscript (HFDS and JBf).
    The inputs undergo a linear projection of the patches to generate the embedded patches. These embeddings are processed by $n_l$ Transformer encoders. Within each encoder block, the data passes through layer normalization (Norm) and a Multi-Head Factored Attention mechanism. This is followed by a Multi-Layer Perceptron (MLP) comprising Dense layers of dimensions $d$ and $d'$, with a GELU activation function. Residual connections (+)  are also included. The output of the encoder blocks is routed through a Dense layer that fist expand the input and then perform a pooling operation. The final outputs are the updated Backflow orbitals or the modified hidden orbitals.}
    \label{fig:vit}
\end{figure*}

\noindent\textbf{Hidden Fermion Determinant States:} 
Instead of adding a many-body correction to non-interacting orbitals, hidden fermion determinant states (HFDS) augment the Fock space of $N$ fermions with $M$ modes by adding $\tilde N$ hidden fermions that occupy $\tilde M$ hidden modes. 
The physical wavefunction amplitudes $\psi(\vb*n)$ are then obtained by projecting the state in the augmented space $\ket{\Phi}$ assuming that the occupancies of the hidden fermions depend on those of the physical ones. 
In this work, we consider the case where $\ket{\Phi}$ is a Slater determinant. 
Therefore, the physical wavefunction is expressed as
\begin{equation}
    \psi(\vb*n) = \prod_\sigma \det\left([\vb*\Phi^{\sigma,(0)}]_{\vb*n} + \tilde{\vb*\Phi}^\sigma(\vb*n) \right),
\end{equation}
where $[\vb*\Phi^{\sigma,(0)}]_{\vb*n}$ are $(N_\sigma + \tilde N) \times (N_\sigma + \tilde N)$ matrices whose last $\tilde N$ rows are zero. 
They are obtained by slicing the $N_\sigma$ rows of the input-independent $M \times( N_\sigma+ \tilde N)$ \textit{visible} matrices that correspond to the occupations of $\ket{\vb*n}$. 
The $\tilde{\vb*\Phi}^\sigma(\vb*n)$ terms are $(N_\sigma + \tilde N) \times (N_\sigma + \tilde N)$ matrices whose first $N_\sigma$ rows are zero. 
They are the output of a neural network evaluated on $\ket{\vb*n}$.
Additional details regarding the HFDS can be found in appendix \ref{appendix:HFDS}.

\subsection{Vision Transformer (ViT)}

Both ans\"atze discussed in the previous section rely on a neural network to parameterize the configuration-dependent matrix $\tilde{\vb*{\Phi}}(\vb*{n})$.
Originally introduced for sequence modelling~\cite{vaswani2017attention}, transformers, later extended to computer vision (vision transformers)~\cite{dosovitskiy2020image}, have been successfully applied as NQS for quantum spin systems~\cite{viteritti2023transformer,viteritti2025}, Rydberg atom arrays~\cite{sprague2024variational} and fermionic systems~\cite{gu2025solving, geier2025attention}.
Here, we apply the vision transformer (ViT) from Ref.~\cite{viteritti2023transformer} as an NQS by using it to parameterize $\tilde{\vb*{\Phi}}(\vb*{n})$ for both ans\"atze.
The ViT is made up of three stages: an embedding, encoder and readout.

Configurations from the computational basis of fermionic systems are numerically encoded as $N_o \times (2S+1)$ bit strings \cite{jafari2008introduction}, where $N_o$ is the number of spatial orbitals and $S$ is the spin. 
Each bit of this string specifies the binary occupation of a spin-orbital.
Making a connection with computer vision, we interpret the orbitals as the spatial degree of freedom of an image and the different spin values as different channels (features) of the same image.
We use this channel-based representation as the starting point for evaluating $\tilde{\vb*{\Phi}}(\vb*{n})$.

First, the input is partitioned into patches~\cite{dosovitskiy2020image,viteritti2023transformer}, which both lowers the computational cost of evaluating the network and acts as a form of inductive bias~\cite{nutakki2025design}.
As illustrated in Fig.~\ref{fig:vit}, the spin-channel representation of $\vb*n$ is split into $N_p$ patches of size $p = b\times b,$ where $b$ is the side length of the patch.
Together with the spin channels this results in $N_p$ vectors of size $2p$, which are then embedded into a higher-dimensional space of dimension $d$.

After the embedding stage, the input is processed through an encoder which employs a factored attention mechanism~\cite{viteritti2023transformer, rende2025queries}, where the input-dependent attention matrix is replaced by a fixed, learnable parameter matrix $\vb J$.
The result is a $d$-dimensional correlated representation that incorporates global contextual information across patches. 
Attention heads $h$ are used to introduce parameter-sharing in $\vb J$, and the encoding block is repeated $n_l$ times.

Finally, the output is pooled over the patch dimension, resulting in a $d$-dimensional representation. 
This is passed through a dense layer with $d'$ hidden units and a GELU activation. 
A final dense layer followed by a tanh activation produces the matrices $\tilde{\vb*\Phi}(\vb*n)$.

This fundamental structure is shared by both the HFDS-ViT and JBf-ViT architectures, which can be thought of as functions designed to generate antisymmetric outputs. 
The primary distinction between these two models lies in their readout shape. 
The HFDS-ViT readout is a set of hidden matrices with shape $\tilde N \times (N_\sigma + \tilde N)$ whilst the JBf readout are $M\times N_\sigma$ matrices. 
This leads to a crucial difference in computational efficiency. 
On one hand, when $\tilde N$ is fixed, the HFDS-ViT scales linearly with the system size. On the other, the JBf-ViT scales quadratically.

\begin{figure*}[t]
    \centering
    \includegraphics[width=\linewidth]{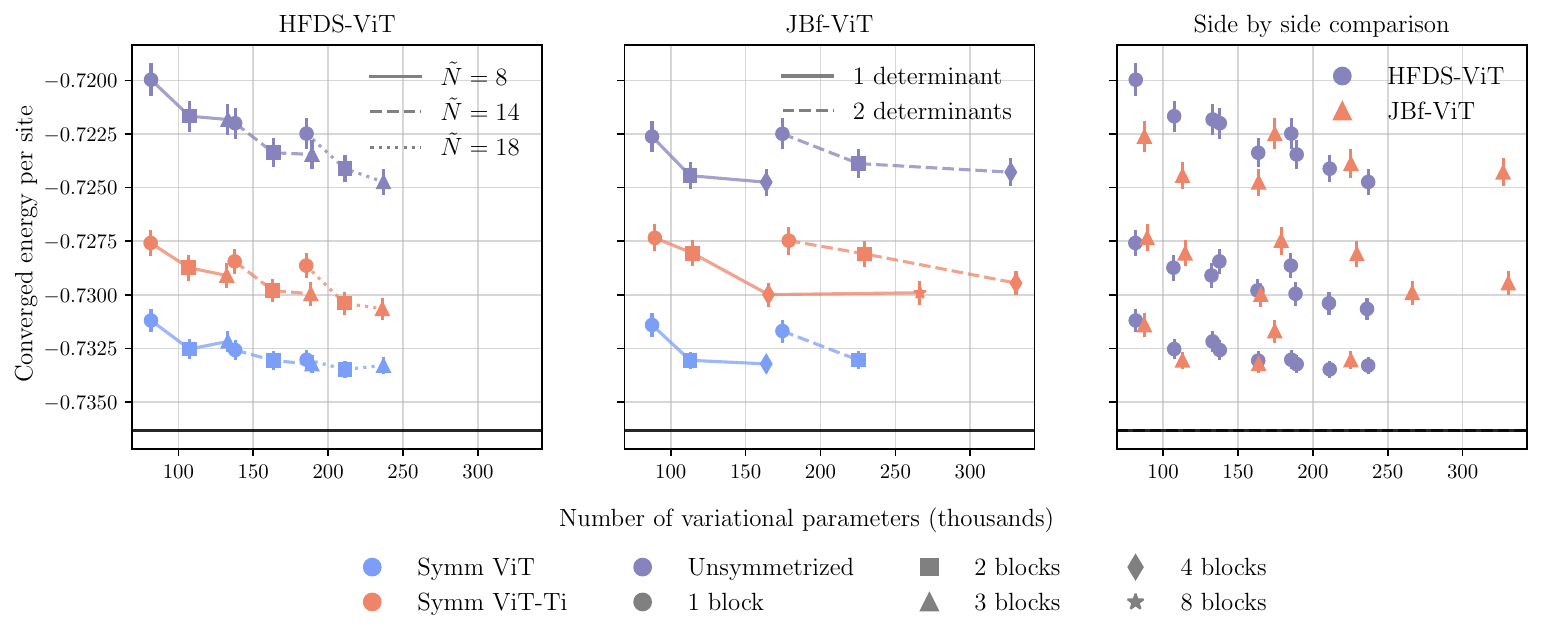}
    \caption{Comparison of variational energies on a the $8 \times 4$ cylinder. The plot shows converged energies for both architectures as a function of the number of variational parameters and symmetrization methods. 
    For HFDS-ViT (left), the number of hidden fermions $\tilde{N}$ is varied. For JBf-ViT (middle), the number of determinants is varied. Symbols indicate different numbers of stacked encoder blocks, $n_l$, in the ViT. The right panel provides a side by side comparison of both methods. The solid black line represents the energy obtained for this system from a DMRG simulation using a bond dimension of 8000 with maximum truncation error of $\simeq1\times10^{-5}$.}
    \label{fig:8x4_symms_and_params_compare}
\end{figure*}
\subsection{Symmetries}
When the Hamiltonian commutes with a set of operators that obey a group structure, it can be block diagonalized \cite{tinkham2003group}.  
Each block corresponds to a particular irreducible representation (irrep) of the group. 
Consequently, the eigenstates of the Hamiltonian can be classified by their irrep~\cite{roth2023high}. 
It has been shown that enforcing Hamiltonian symmetries on a variational wavefunction can significantly improve variational energies~\cite{tahara2008, morita2015, hibat-allah2020, nomura2021helping, robledo2022fermionic, reh2023, medvidovic2024neural}.
In general, an unrestricted wavefunction will not exactly belong to a specific irrep after optimization. This symmetry may be restored by using quantum number projection~\cite{mizusaki2004, nomura2021helping}.
This is done using
\begin{equation}
    \psi_\mu(\vb*n) = \frac{1}{|G|} \sum_{g \in G} \chi_\mu^\ast(g) \xi_{g^{-1}}(\vb*n) \psi(\vb*n \circ g), 
    \label{eq:symmetry_projection}
\end{equation}
where $\mu$ is the index of the irreducible representation, $|G|$ is the order of the group, $\chi_\mu(g)$ is the character of the irreducible representation evaluated on $g$, $\xi_g(\vb*n)$ is the sign of the permutation applied to state $\ket{\vb*n}$, and $\vb*n \circ g$ are the permuted occupation numbers. More details concerning the representation of permutation groups on fermionic Fock space are given in appendix~\ref{appendix:symmetries}.

\Cref{eq:symmetry_projection} can be implemented using a brute-force $O(\abs{G})$ approach, with every amplitude query $\psi_\mu(\vb*{n})$ requiring  $|G|$ neural-network evaluations.
However, when the order of the group scales with the lattice size, as is the case for the translation group, this leads to an $O(M)$ overhead which can be prohibitively expensive for large lattices.

To avoid multiple network evaluations, it is possible to design attention matrices that produce patch-translation equivariant attention vectors~\cite{viteritti2023transformer}\footnote{While it would be desiderable to have the neural network output be fully translational-equivariant, after the patching it is only possible to maintain equivariance to translations that send patches into other patches.}. 
To do so, the attention weights need to satisfy $J_{ij}^\mu = J^\mu_{T(i) T(j)}$, a derivation of this condition can be found in \Cref{appendix:equivariant_attention}.

When the output of the last encoder block is summed over the patch dimension, this results in a patch-translation-invariant output. 
In order to obtain an output that is invariant with respect to the full translation group of the lattice, the output is pooled over the intra-patch translation vectors $\vb*p$,
\begin{equation}
    \tilde{\vb*\Phi}(\vb*n) = \sum_{\vb*p} f_{\text{ViT}}(\vb*n \circ T_{\vb*p}).
\end{equation}
Here, we consider a square lattice which is partitioned into $2\times2$ patches, so these vectors are $\vb*p \in \{\vb*0, \vb*a_1, \vb*a_2, \vb*a_1+\vb*a_2\}$ where $\vb*a_i$ are the primitive lattice vectors. This construction leads to $\tilde{\vb*\Phi}(\vb*n) = \tilde {\vb*\Phi}(\vb*n \circ T_{\vb*R})$ for all lattice vectors $\vb*R$. Consequently, a translationally-symmetrized wavefunction with momentum $\vb*k$ is given by
\begin{align}
    \psi_{\vb*k}(\vb*n) = \frac{1}{M} \sum_{\vb*R} e^{-i \vb*k \cdot \vb*R} \xi_{-\vb*R}(\vb*n) \nonumber\\
    \times \det \left([\vb*\Phi^{(0)}]_{\vb*n \circ T_{\vb*R}} + \tilde{\vb*\Phi}(\vb*n) \right),
    \label{eq: ViT_Ti}
\end{align}
where $\vb*k$ is a reciprocal vector in the first Brillouin zone. Equation \ref{eq: ViT_Ti} produces translationally-invariant amplitudes, in which the ViT only needs to be evaluated on the vectors $\vb*p$~\cite{romero2025}. The number of such vectors remains constant with the size of the system. 

We denote the translationally invariant wavefunction produced from equation~\ref{eq: ViT_Ti} as \textbf{Symm ViT-Ti} and the one produced from equation \ref{eq:symmetry_projection} as \textbf{Symm ViT.}

\section{Results}
\label{sec:numerical_results}
In this section, we present results obtained using the symmetrization procedures described above, applied to both the HFDS and JBf wavefunctions, each parametrized by a ViT. 

Note that while the JBf explicitly incorporates a density-density Jastrow factor, we intentionally omit an explicit Jastrow factor from the HFDS ansatz as it is theoretically and empirically redundant to accuracy (see Appendix \ref{appedix:jastrow_hfds}).
We focus on the $8 \times L$ square lattice where $L \in \{4,8\}$ using both open and periodic boundary conditions. 
In the simulations, we fix $d=64$, $h=8$ and $d'=128$. The wavefunctions are optimized using the kernel reformulation~\cite{rende2024simple,Chen2024} of stochastic reconfiguration~\cite{PhysRevB.71.241103}.
The learning rate is gradually annealed using a cosine decay scheduler \cite{loshchilov2016sgdr} whilst the diagonal shift of the neural tangent kernel is linearly decreased from $10^{-2}$ to $10^{-8}.$
We do not use a pinning field and start from a set of random variational parameters.

\subsection{$8 \times 4$ cylinder (OBC-PBC)}
\label{sec:cylinder_numerical_results}
First, we consider an $8 \times 4$ lattice with open boundary conditions along the long side and periodic boundary conditions along the short side. 

To investigate the effect of the number of parameters in the wavefunction, we systematically increased the number of hidden fermions $\tilde N$ for the HFDS ansätze, the number of determinants for the JBf wavefunction and the number of encoder blocks for both architectures, as shown in~\Cref{fig:8x4_symms_and_params_compare}.
Increasing the number of parameters by increasing the number of encoder blocks tends to improve the variational energy, although the improvement eventually saturates.
While increasing the number of hidden fermions in the HFDS also tends to improve variational energies, this is not necessarily the case when using a second determinant in the JBf. The observed saturation at high parameter count can be attributed to the optimization landscape becoming more challenging to navigate, rendering the ansatz prone to getting trapped in poor local minima. This issue has already been observed and discussed in the setting of fermionic NQS \cite{loehr2025enhancing}.
In what follows we fix $n_l = 2$, $\tilde N=18$ and use a single-determinant in the JBf.

To investigate the accuracy gains of the symmetrization procedure, we also compare both symmetrized and unsymmetrized versions of the ansätze in~\Cref{fig:8x4_symms_and_params_compare}.
The symmetrized wavefunctions are projected onto irreps of the space group and spin-parity group, such that $|G|=32$.
Note that in the case of the Symm ViT-Ti, one can construct a wavefunction that is invariant with respect to the space group whilst only evaluating the network on the point group, whose order does not scale with the system size.
For both the JBf and HFDS ansätze, we find that the lowest energy state corresponds to the same non-trivial irrep of the space group with momentum $k_y = \pi$. 

We find that symmetrized wavefunctions significantly outperform the non symmetrized ones in terms of energy, with the Symm-ViT obtaining the best results. 
This suggests that enforcing symmetry through output averaging, rather than by restricting the model's internal parameters is more expressive, consistent with findings in~\cite{reh2023}.

We also note that on this system, our methods do not outperform DMRG, our lowest energy being 0.4$\%$ higher than the DMRG energy.
In our simulations, all wavefunctions required approximately the same number of steps to reach energy convergence ($\approx 5000$). Beyond the number of variational parameters and optimization steps, the performance and computational cost is impacted by the number of network evaluations per step, which is proportional to the number of Monte Carlo samples being used, and may increase with certain symmetrization strategies.

\begin{figure}[t]
    \centering
    \includegraphics[width=0.9\columnwidth]{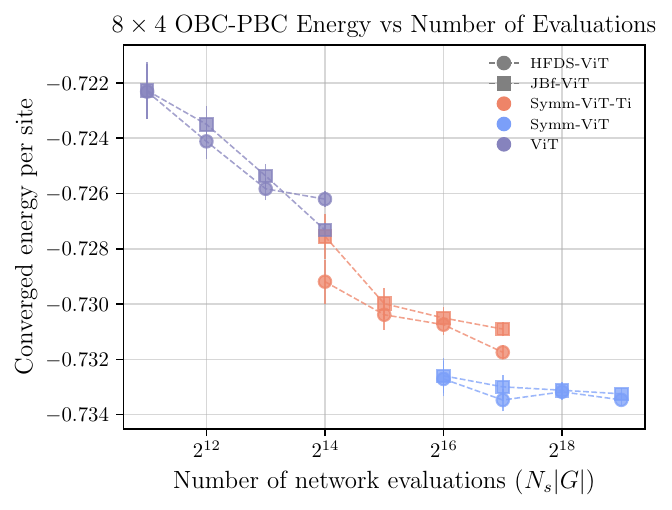}
    \caption{
    Energy per site vs number of network evaluations defined as $N_{s} \times |G|$ for both HFDS-ViT and JBf-ViT for the three symmetrization methods on the $8 \times 4$ cylinder. The size of $|G|$ depends on the symmetrization technique used. $|G|=1$ for ViT, $|G|=8$ for Symm-ViT-Ti and $|G|=32$ for Symm-ViT.}
    \label{fig:8x4_symms_and_samples_compare}
\end{figure}

We also investigate how increasing the number of Monte Carlo samples impacts the accuracy of the converged energy. 
Specifically, we select the best hyperparameters identified in the previous analysis and re-optimized them using 2048, 4096, 8192 and 16384 samples. 
The results are presented in Fig. \ref{fig:8x4_symms_and_samples_compare}. Across all architectures, increasing the number of samples consistently improves the final energy. However, the magnitude of this improvement depends strongly on the ansatz. In particular, the non-symmetrized ViT exhibits the most pronounced improvement as the number of samples increases, suggesting that its optimization is more sensitive to Monte Carlo noise. In contrast, the symmetrized ansätze show a weaker dependence on the number of samples.

Interestingly, in Fig.\ref{fig:8x4_symms_and_samples_compare}, we observe an approximate scaling of the accuracy with the total number of network evaluations, which corresponds to the number of samples $N_s$ times the size of the group with respect to which the wavefunction is explicitly projected (see Eq.~\eqref{eq:symmetry_projection}).

This suggests that more network evaluations, whether they come from increasing the number of Monte Carlo samples or from enforcing symmetries, lead to increased accuracy. 
However, as the optimization algorithm (minSR) requires the inversion of an $N_s \times N_s$ matrix, the most efficient scheme will maximize the wavefunction evaluations while minimizing $N_s$.
This is consistent with our empirical observation.

\begin{figure*}[t]
    \centering
    \includegraphics[width=\linewidth]{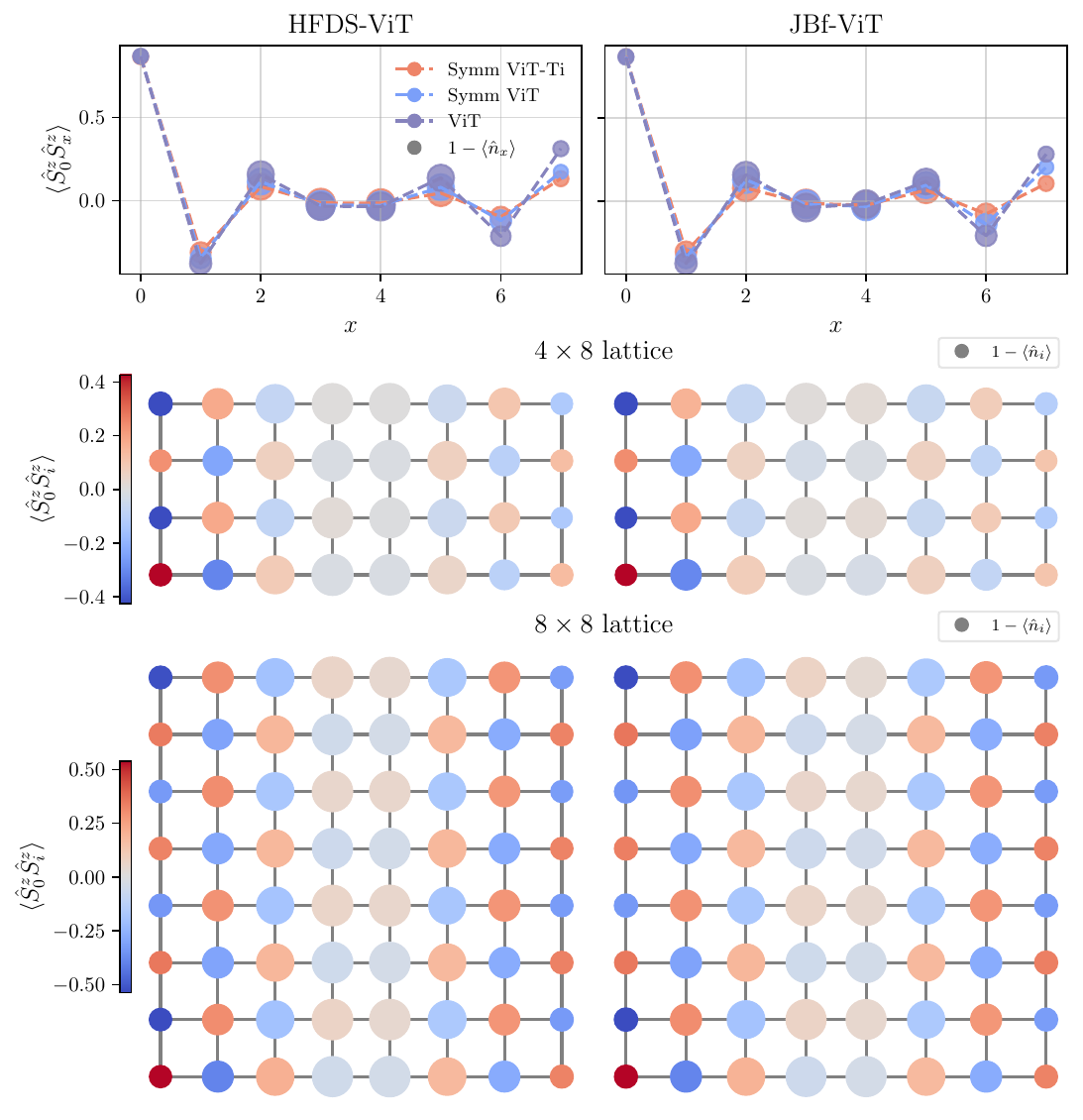}
    \caption{Comparison of physical observables on the $8 \times 4$ and $8 \times 8$  cylinders. The first row shows a horizontal cut of the spin-spin correlation. The $x$ axis corresponds to the site index and the size of the balls is proportional to the hole density at that site. The central (bottom) row plots show the observables in real space for the lowest energy solution on the $8 \times 4$ ($8 \times 8$) system.}
    \label{fig:8x4_observables}
\end{figure*}

To investigate the presence of charge and spin orders, we compute the spin-spin correlation function $S_{ij} = \expval*{\hat S_i^z \hat S_j^z}$ and the hole density $h_i = 1-\expval{\hat n_i},$ where the expectation value is taken with respect to the ground state and computed stochastically~\cite{binder2012monte}.
Plots of these observables for both the JBf and HFDS are presented in Figure~\ref{fig:8x4_observables}.

The spin–spin correlation function and hole density exhibit a consistent pattern across all three approaches. 
For both the HFDS and JBf ansätze, we observe regions of high hole concentration that align with domain walls in the antiferromagnetic order. 
This is consistent with the $\lambda = 8$ filled stripe discussed in~\Cref{sec:Hubbard_model}.
While the overall structure of the correlations is similar across the different architectures, we note small differences in the amplitude of the spin–spin signal and the hole density, reflecting subtle variations in how each network captures charge and magnetic ordering.

\subsection{8x8 torus (PBC-PBC)}
\label{sec:torus_numerical_results}
We now turn to the 
$8 \times 8$ square lattice, with periodic boundary conditions in both directions.
In this case, we fix the network hyperparameters to $d=64$, $h=8$, $n_l=2$, $4096$ samples, $\tilde N = 22$ and one determinant in the JBf. For the HFDS, we use $d'=128$ whereas for the JBf the latent dimension of the readout is set $d'=64$, which results in a similar number of parameters.
Due to the favorable scaling of the HFDS, we can use a higher $d'$ for the same number of parameters.
First, several identical Symm-ViT wavefunctions are initialized and optimized with respect to the translation group for the $\vb*k$ points of the irreducible Brillouin zone.
After identifying the lowest energy $\vb*k$-point, the optimization is continued in that momentum sector until convergence.
For both the JBf-ViT and HFDS-ViT ansätze, we found that $\vb*k=(\pi/2, \pi/2)$ produces the lowest energy.

To study the magnetic and electronic properties of  the converged solution, we compute the spin-spin correlation, the density-density correlation,
\begin{equation}
    \rho_{ij} = \expval*{\hat n_i \hat n_j} - \expval{\hat n_i} \expval{\hat n_j}
\end{equation}
as well as the associated structure factors. The structure factor $\tilde A_{\vb*k}$ of a translationally-invariant two-point correlation function $A_{ij}$ is given by
\begin{equation}
    \tilde  A_{\vb*k} = \sum_{i} e^{i \vb*k \cdot (\vb*R_i - \vb*R_j)} A_{ij},
\end{equation}
where $\vb*R_i$ is position of site $i$ and $\vb*k$ a vector in the first Brillouin zone. 
\begin{figure*}[th!]
    \centering
    \includegraphics[width=\linewidth]{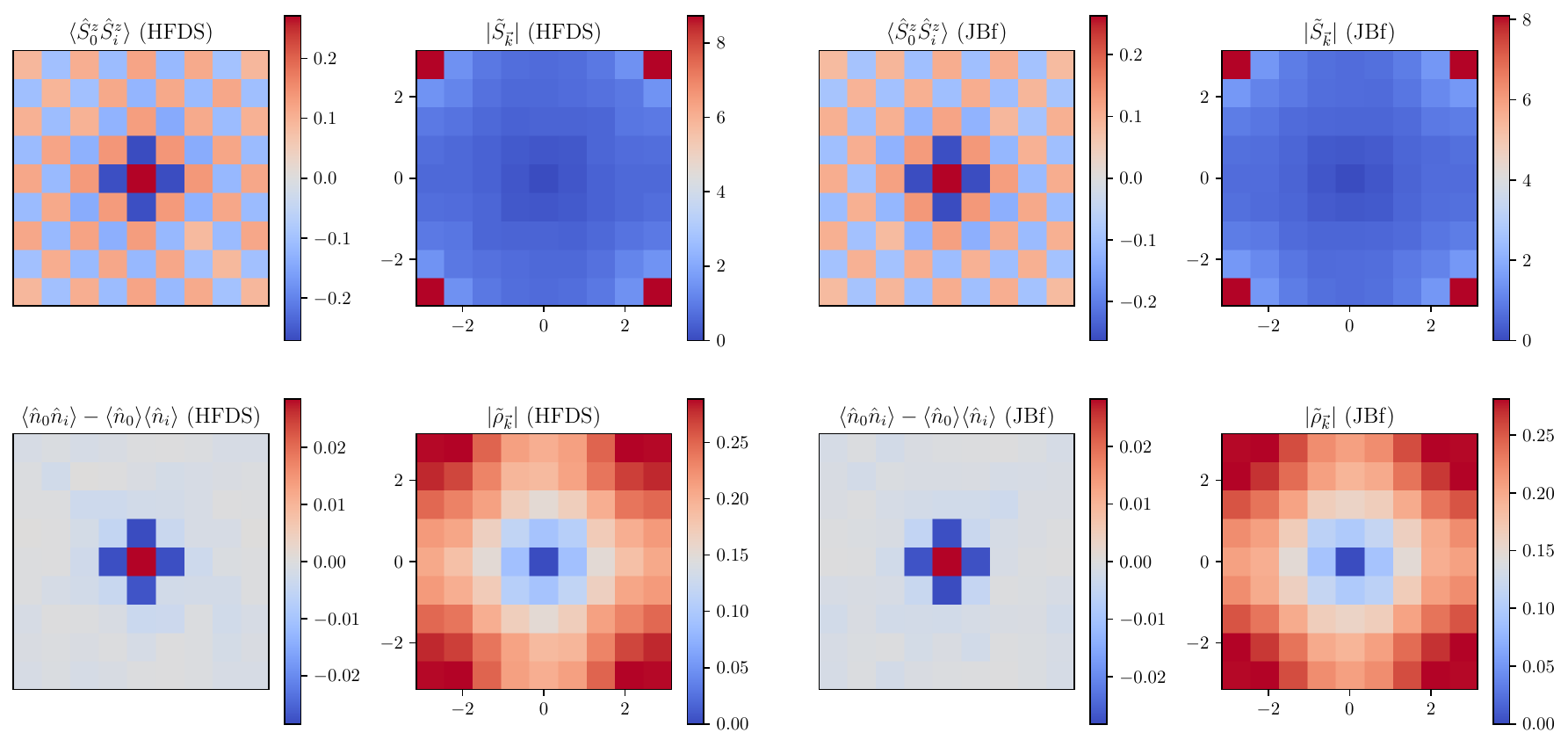}
    \caption{Comparison of physical observables on the $8 \times 8$ square lattice (PBC-PBC) with $U/t=8$ and $n_h= 1/8$ hole doping. The results show real space plots of the expectation values of the of spin-spin correlation function, the density-density correlation and the associated structure factors for both ansätze. The expectation values are computed from the converged Symm-ViT variational states, the symmetrization being with respect to the translation group with momentum $\vb*k = (\pi/2, \pi/2)$, previously identified as the lowest energy sector.}
    \label{fig:8x8observables}
\end{figure*}

Both methods display a similar charge and spin ordering pattern (see~Fig~\ref{fig:8x8observables}). The spin-spin signal exhibits long-range antiferromagnetic correlations, characterized by sharp peaks at $\vb*k=(\pi,\pi)$ in the structure factor.
In contrast, the density-density correlations are short ranged and uniform, indicating suppressed density fluctuations. These results are consistent with the physics of a doped Mott insulator. This state is distinct from the $\lambda = 8 $ filled stripe phase, as the required $2\lambda = 16$ spin density periodicity is incompatible with the $8 \times 8$ torus.

A similar calculation was performed on this system but with open boundary conditions in one direction. The results on Fig. \ref{fig:8x4_observables} show that $\lambda=8$ stripes are found.

We also explored the performance of the ViT and ViT-Ti architectures on this larger system. 
Using the same neural network hyperparameters, we trained all models to convergence. 
For ViT-Ti, we focused on the momentum sector previously identified as energetically favorable, and compared the final energies as well as the physical observables. Surprisingly, the unsymmetrized wavefunction outperforms the ViT-Ti, obtaining an energy lower by about $1.2\%$ see \ref{tab:combined_energy}.
A possible explanation for this discrepancy is that enforcing full translational equivariance on the attention weights imposes an overly restrictive constraint on the variational wavefunction.
In addition, the unsymmetrized wavefunctions struggle to capture the spin-spin correlation function.

\section{Conclusion and outlook}
\label{sec:discussion_conlusion}
In this work, we used the ViT architecture to represent the backflow orbitals of the JBf wavefunction and the hidden orbitals of the HFDS. Our simulations of the $8 \times L$ Hubbard model, with $L\in \{4,8\}$, in the strongly correlated, underdoped regime, provide a systematic comparison between these two ans\"atze.

Our results show that HFDS and JBf wavefunctions achieve comparable variational energies and observables in practice.
Although previous studies reported differing performance between the two, we attribute these discrepancies to differences in network architecture, size, and training protocols~\cite{luo2019backflow, robledo2022fermionic}, in line with the findings of Ref.~\cite{liu2024unifying}.
A key distinction arises in scalability: HFDS scales more favorably with system size, as its readout depends on a fixed number of hidden fermions, potentially providing an advantage for simulating larger systems.

We find that enforcing symmetries of the Hamiltonian is crucial in order to obtain lower energies and physically-meaningful observables.
Our results show that explicitly imposing symmetries through computationally expensive output averaging outperforms architectures that enforce constraints more efficiently \cite{romero2025}.

Finally, both methods converged to the expected $\lambda = 1/n_h$ filled stripes on cylindrical systems, in line with recent results in the literature \cite{qin2022, liu2025accurate, gu2025solving}. 
On the $8 \times 8$ torus, we find physics which is characteristic of a doped Mott-insulator, marked by long range antiferromagnetic correlations and localized, repulsive density fluctuations. 
We are however not able to access system sizes large enough to resolve the presence of incommensurate spin and charge ordering. Therefore, our results may be impacted by significant finite-size effects. 

This highlights a crucial challenge for future research, developing methods that incorporate symmetries in an effective and scalable manner. 
Overcoming this hurdle is essential for probing larger system sizes, in order to obtain better resolution of ground state properties and reliably extrapolating to the thermodynamic limit.

Our results suggest that a path forward for efficiently simulating larger systems could be achieved by using the HFDS with a reduced number of hidden fermions allowing one to adopt low-rank determinant updates in the spirit of Ref.~\cite{chen2025neuralnetworkaugmentedpfaffianwavefunctions}.Furthermore, while it is unknown whether the comparable performance of these architectures extends to more complicated models, such as the extended Hubbard model or systems with disordered interactions ($U_i, t_i$), exploring these generalizations serves as a promising future research direction.
\section{Acknowledgements}
    Simulations were performed with NetKet~\cite{netket2:2019,vicentini2022netket}, and at times parallelized with mpi4JAX~\cite{mpi4jax:2021}. 
    This software is built on top of JAX \cite{jax2018github} and Flax \cite{flax2020github}.
    We acknowledge insightful discussions with G. Carleo and A. Kahn. 
    F.V. acknowledges support by the French Agence Nationale de la Recherche through the NDQM project ANR-23-CE30-0018. O.S. acknowledges support from the Swiss National Science Foundation through the Postdoc.Mobility fellowship.
    This project was provided with computing HPC and storage resources by GENCI at IDRIS thanks to the grant 2024-A0170515698 on the supercomputer Jean Zay's A100 partition.
    We acknowledge EuroHPC Joint Undertaking for awarding us access to Leonardo at CINECA, Italy through grant EHPC-AI-2024A05-006 .
\bibliography{./bibliography}
\clearpage
\appendix
\onecolumngrid

\section{Hidden fermion determinant state}
\label{appendix:HFDS}
The Fock space of a spin $\frac 12$ fermionic system on a finite lattice with $M$ sites is given by all the ways that the $2M$ single-particle states can be occupied:
\begin{equation}
    \ket{\vb*n} = \prod_{i, \sigma} \hat c^{\dag n_{i\sigma}}_{i\sigma}\ket{\varnothing},
\end{equation}
where $\ket{\varnothing}$ denotes the Fock vacuum and $n_{i\sigma} \in \{0,1\}$ due to the Pauli exclusion principle. In the Hidden Fermion formalism, we add $\tilde N$ hidden fermions that can occupy $\tilde M$ hidden modes. The states of this space are now also specified by the occupations of the hidden orbitals:
\begin{equation}
    \ket{\vb*n, \tilde {\vb*n}} = \prod_{i,\sigma} \hat c^{\dag n_{i\sigma}}_{i\sigma} \prod_{\nu }\hat d^{\dag \tilde n_\nu}_\nu \ket{\varnothing}.
\end{equation}
The occupations of the hidden modes depend on those of the visible ones through a constraint function: $\tilde{\vb*n} = F(\vb*n),$ such that the amplitudes of the target wavefunction in the Fock basis are obtained by taking the following overlap
\begin{equation}
    \psi(\vb*n) = \braket{\vb*n, F(\vb*n)}{\Phi},
\end{equation}
where $\ket{\Phi}$ is a state of the augmented space. We consider the case in which $\ket{\Phi}$ is a Slater determinant. This means that we fix $\tilde N$ and work in sectors of the physical space that have a fixed number of particles and magnetization. The Slater determinant can be written down in second-quantized form as:
\begin{equation}
    \ket{\Phi} = \prod_{\alpha=1}^{N_{\text{tot}}} \hat \varphi^\dag_\alpha \ket{\varnothing},
\end{equation}
where $N_\text{tot} = N_\up+\tilde N + N_\down + \tilde N$ and the operators $\hat \varphi^\dag_\alpha$ are linear combinations of the visible and hidden operators. All of this information can be stored into an $M_\text{tot} \times N_{\text{tot}}$ Slater matrix $\vb*\Phi$, which we choose to be block diagonal in the spin basis. It is convenient to represent each Slater matrix in the following block form
\begin{equation}
    \vb*\Phi = \begin{pmatrix}
        \vb*\Phi^\down & 0 \\ 0 & \vb*\Phi^\up
    \end{pmatrix}, \quad \vb*\Phi^{\sigma}  = \vb*\Phi^{\sigma, (0)} +  \vb*{\tilde\Phi}^\sigma
\end{equation}
where $\vb*\Phi^{\sigma, (0)}$ is the $(M+\tilde M) \times(N_\sigma+\tilde N)$ \textit{visible} matrix whose last $\tilde M$ rows are zero and $\vb*{\tilde \Phi}^\sigma$ is the $(M+\tilde M) \times(N_\sigma+\tilde N)$ \textit{hidden} matrix whose first $M$ rows are set to zero. Given a basis state $\ket{\vb*n},$ the wavefunction is given as: 
\begin{equation}
    \psi(\vb*n) = \det \vb*\Phi(\vb*n) = \prod_\sigma \det\left( [\vb*\Phi^{\sigma, (0)}]_{\vb*n} + \vb*{\tilde \Phi}^\sigma(\vb*n) \right),
\end{equation}
where $\vb*\Phi^{\sigma, (0)}(\vb*n)$ (resp. $\vb*{\tilde \Phi}^\sigma(\vb*n)$) is obtained by slicing the rows of the visible (resp. hidden) matrix that correspond to the occupations of $\ket{\vb*n}$ (resp. $F(\vb*n)$). In our implementation, the visible matrices are configuration-independent. The sliced hidden matrix is directly obtained as the output of a neural network.

\section{Symmetries}
\label{appendix:symmetries}
Let $\mathcal F$ be a fermionic Fock space with $m$ single-particle states and $G$ a subgroup of $\mathcal{S}_m$. Elements of $\mathcal F$ are expressed of products of fermionic creation operators acting on the Fock vacuum

\begin{equation}
    \ket{\vb*n}  = \hat c^{\dag n_{\alpha_1}}_{\alpha_1} \hat c^{\dag n_{\alpha_2}}_{\alpha_2} \ldots \hat c^{\dag n_{\alpha_m}}_{\alpha_m} \ket{\varnothing}
\end{equation}
where a canonical ordering has been defined, i.e, $\alpha_1 < \alpha_2 < \ldots < \alpha_m$. In order to define a unitary representation $\hat U$ of $G$ on $\mathcal{F}$ one must specify how the creation operators transform under conjugation and how the representation acts on the Fock vacuum. Consequently, we define
\begin{enumerate}
    \item For all $g \in G$, $\hat U_g \ket{\varnothing} = \ket{\varnothing}$. 
    \item For all $g \in G$ and all single particle states $\alpha$, $\hat U_g \hat c^\dag_\alpha \hat U_g^\dag = \hat c^\dag_{g(\alpha)}$
\end{enumerate}
where $g(\alpha)$ denotes the permuted state index. As a result, the action of the representation on an arbitrary Fock state is expressed as:

\begin{equation}
    \hat U_g \ket{\vb*n} = \xi_g(\vb*n) \ket{\vb*n \circ g^{-1}}.
    \label{eq: fermion_representation_action}
\end{equation}
We then define the symmetrizer projector $\mathcal{\hat P}_\mu$ operator for a particular irreducible representation $\mu$ as:

\begin{equation}
    \mathcal{\hat P}_\mu = \frac{d_\mu}{|G|} \sum_{g \in G} \chi_\mu^\ast(g) \hat U_g.
\end{equation}
where $d_\mu$ is the dimension of the representation, $|G|$ is the group order and $\chi_\mu(g)$ is the character of the irreducible representation evaluated on element $g.$ The symmetrizer can thus be used to construct amplitudes that behave like irreps of $G$:

\begin{equation}
    \psi_\mu(\vb*n) = \mel*{\vb*n}{\mathcal{\hat P}_\mu}{\psi} = \frac{d_\mu}{|G|} \sum_{g \in G} \chi^\ast_{\mu}(g) \mel{\vb*n}{\hat U_g}{\psi} = \frac{d_\mu}{|G|} \sum_{g \in G} \chi^\ast_\mu(g) \xi_{g^{-1}}(\vb*n) \psi(\vb*n \circ g)
\end{equation}
This followed from $\bra{\vb*n} \hat U_g = \left( \hat U_g^\dag \ket{\vb*n}\right)^\dag$ and the fact that $\hat U_g^\dag = \hat U_{g^{-1}}.$ \\

\section{Equivariant attention vectors}
\label{appendix:equivariant_attention}
Suppose we have a permutation group $G$ that permutes the embedded vectors $\vb*x_j$ to $\vb*x_{g(j)}.$ The condition for the attention vectors to be equivariant with respect to all $g$ in $G$ is: 

\begin{equation}
    \vb*A^\mu_{g(i)}(\vb*x) = \vb*A^\mu_i(\vb*x \circ g).
\end{equation}
This means that:
\begin{equation}
    \sum_{j=1}^{N_p} J_{g(i) j}^\mu \vb V^\mu \vb*x_j = \sum_{j=1}^{N_p} J_{ij}^\mu \vb V^\mu \vb*x_{g(j)} = \sum_{j=1}^{N_p} J_{i g^{-1}(j)}^\mu \vb V^\mu \vb*x_j
\end{equation}
where we used a change of index $k = g(j).$ This leads to $J_{g(i) j}^\mu = J_{i g^{-1}(j)}^\mu$ or in other words, $J_{g(i) g(j)}^\mu = J_{ij}^\mu.$ \\ 

\section{Translation invariance of
symm(ViT-Ti)}
\label{appendix:proof_ViT-Ti}

In this section, we show how a translationally invariant state can be used to obtain an invariant wavefunction. Without loss of generality, we present the result for a symmetrized wavefunction in the trivial sector of the translation operator; however, the same reasoning applies to a generic symmetry transformation in any sector.\\
Starting from equation \eqref{eq:symmetry_projection}, consider a translated input.
\begin{equation*}
    \psi_{\text{Sym}}\big( \vb*n \circ T_q \big) = \frac{1}{|G|} \sum_{g} \chi^{*}(g) \, \xi_{g^{-1}}\big(\vb*n \circ T_q \big) \, \psi\big( \vb*n \circ T_g \, T_q\big).
\end{equation*}
Noting that the fermionic sign function transforms as
$
    \xi_{g^{-1}}\big(\vb*n \circ T_q\big) = \xi_{{g^{-1}} \cdot {q^{-1}}}(\vb*n), 
$
and that the translation operation composes as
$
    \psi\big( \vb*n \circ T_g \, T_q\big) = \psi\big( \vb*n \circ T_{g\cdot q}\big),
$
the expression becomes
\begin{equation*}
    \psi_{\text{Sym}}\big(\vb*n \circ T_q \big) = \frac{1}{|G|} \sum_{g}\chi^{*}(g)\, \xi_{{g^{-1}} \cdot {q^{-1}}}(\vb*n) \, \psi\big(\vb*n \circ T_{g\cdot q}\big).
\end{equation*}
By performing a change of variable \(g' = g \cdot q\) (with the inverse transformation \(g = g' \cdot q^{-1}\)), the sum becomes:
\begin{equation*}
    \psi_{\text{Sym}}\big(\vb*n \circ T_q\big) = \frac{1}{|G|} \sum_{g'} \chi^{*}(g'\cdot q^{-1}) \, \xi_{(g^{-1})'}(\vb*n) \, \psi\big(\vb*n \circ T_{g'}\big).
\end{equation*}
Since the characters satisfy the group property
$
    \chi^{*}(g'\cdot q^{-1}) = \chi^{*}(g') \, \chi^{*}(q^{-1}).
$
Therefore, the expression factors
\begin{equation*}
    \psi_{\text{Sym}}\big(\vb*n \circ T_q\big) = \chi^{*}(q^{-1}) \left[\frac{1}{|G|} \sum_{g'} \chi^{*}(g') \, \xi_{(g^{-1})'}(\vb*n) \, \psi\big(\vb*n \circ T_{g'}\big)\right].
\end{equation*}
Recognizing that the term in brackets is exactly \(\psi_{\text{Sym}}(\vb*n)\), we obtain:
\begin{equation}
    \psi_{\text{Sym}}\big(\vb*n \circ T_q\big) = \chi^{*}(q^{-1}) \, \psi_{\text{Sym}}(\vb*n).
\label{symm Ti-ViT:symmeq}
\end{equation}
This shows that the symmetric state \(\psi_{\text{Sym}}(\vb*n)\) transforms according to the character \(\chi^{*}(q^{-1})\) under any translation \(T_q\), thereby preserving the invariance and antisymmetry of the wavefunction.\\
Now, we can reach the same result using a translational invariant network for the backflow transformation. Here, we need to be more explicit with the notation.\\
The wavefunction is defined as follows:
\begin{equation*}
    \psi_{\text{Sym}}(\vb*n) = \frac{1}{|G|} \sum_{g} \chi^{*}(g) \xi_{g^{-1}}(\vb*n) \cdot J(\vb*n) \prod_{\sigma} \det\Big( [\vb*\Phi^{\sigma,(0)}+\tilde{\vb*\Phi}^\sigma(\vb*n)]_{\vb*n \circ T_{g}} \Big),
\end{equation*}
where \(J(\vb*n)\) is the Jastrow factor, \(\vb*\Phi^{\sigma, (0)}\) denotes the mean-field orbitals matrix, and \(\tilde{\vb*\Phi}^\sigma(\vb*n)\) represents the backflow transformation matrix.

The subscript in 
$
[\vb*\Phi^{\sigma,(0)}+\tilde{\vb*\Phi}^\sigma(\vb*n)]_{\vb*n \circ T_{g}}
$
indicates that, given the full matrix \(\vb*\Phi^{\sigma,(0)}+\tilde{\vb*\Phi}^\sigma(\vb*n)\), we select only the occupied orbitals corresponding to the configuration string \(\vb*n\).\\
Hence, given a translation of the input by \(T_q\), the wavefunction becomes:
\begin{equation*}
    \psi_{\text{Sym}}(\vb*n \circ T_{q}) = \frac{1}{|G|} \sum_{g} \chi^{*}(g) \, \xi_{g^{-1}\cdot q^{-1}}(\vb*n) \cdot J\big(\vb*n \circ T_{q}\big) \prod_{\sigma}\det\Big( [\vb*\Phi^{\sigma, (0)}+\tilde{\vb*\Phi}^\sigma(\vb*n \circ T_{q})]_{\vb*n \circ T_{g\cdot q}} \Big).
\end{equation*}
But now, assuming that both the backflow transformation and the Jastrow factor yield invariant representations, we have:
\[
J\big(\vb*n \circ T_{q}\big)= J(\vb*n) \quad \text{and} \quad \tilde{\vb*\Phi}^\sigma(\vb*n \circ T_{q})= \tilde{\vb*\Phi}^\sigma(\vb*n).
\]
Thus, we can rewrite the expression as:
\begin{equation*}
    \psi_{\text{Sym}}(\vb*n) = \frac{1}{|G|} \sum_{g'} \chi^{*}(g' \cdot q^{-1})\xi_{g'}(\vb*n) \cdot J(\vb*n) \prod_{\sigma} \det\Big( [ \vb*\Phi^{\sigma,(0)}+\tilde{\vb*\Phi}^\sigma(\vb*n)]_{\vb*n \circ T_{g'}} \Big),
\end{equation*}
where we have re-indexed the sum using \(g' = g\cdot q\).\\
Since, again, the characters satisfy the group property
$
    \chi^{*}(g' \cdot q^{-1}) = \chi^{*}(g') \, \chi^{*}(q^{-1}),
$
the expression factors

\begin{equation*}
    \psi_{\text{Sym}}\big(\vb*n \circ T_q\big) = \chi^{*}(q^{-1}) \bigg\{\frac{1}{|G|} \sum_{g'} \chi^{*}(g')\xi_{(g')^{-1}}(\vb*n) \cdot J(\vb*n) \prod_{\sigma} \det\Big( [\vb*\Phi^{\sigma,(0)}+\tilde{\vb*\Phi}^\sigma(\vb*n)]_{\vb*n \circ T_{g'}} \Big) \bigg\}.
\end{equation*}
Recognizing that the term in brackets is exactly \(\psi_{\text{Sym}}(\vb*n)\), we obtain:
\begin{equation*}
    \psi_{\text{Sym}}\big(\vb*n \circ T_q\big) = \chi^{*}(q^{-1}) \, \psi_{\text{Sym}}(\vb*n).
\end{equation*}
Thus, we retrieve exactly the same equation \eqref{symm Ti-ViT:symmeq} without needing to evaluate the network \(|G|\) times.
\section{Variational energies}
\begin{table}[!h]
\centering
\begin{minipage}{0.48\textwidth}
\centering
\begin{tabular}{|l|c|c|}
\hline
Method & $E_0/M$ & $\sigma^2/M$ \\
\hline 
DMRG ($8 \times 4$) &  -0.736329(9) & NA \\
HFDS ($8 \times 4$) & -0.73342(8) & 0.0225 \\
HFDS ($8 \times 8$ Torus) & -0.7454(9) & 0.0563 \\
HFDS ($8 \times 8$ Cylinder) & -0.7259(0) & 0.0422 \\
JBf ($8 \times 4$) & -0.7332(6) & 0.0235 \\
JBf ($8 \times 8$ Torus) & -0.7458(6) & 0.0525 \\
JBf ($8 \times 8$ Cylinder) & -0.7264(5) & 0.0334 \\
\hline 
\end{tabular}
\end{minipage}%
\hfill
\begin{minipage}{0.48\textwidth}
\centering
\begin{tabular}{|l|c|c|}
\hline
Method & $E_0/M$ & $\sigma^2/M$ \\
\hline
HFDS-ViT & -0.736(1) & 0.105 \\
HFDS Symm-ViT-Ti & -0.729(4) & 0.139 \\
JBf-ViT  & -0.7422(3) & 0.0698 \\
JBf Symm-ViT-Ti & -0.726(9) & 0.168 \\
\hline
\end{tabular}
\end{minipage}
\caption{Converged variational energies per site for different lattices and network architectures. The left table shows results for the $8 \times 4$ and $8 \times 8$ cylinders and $8 \times 8$ torus lattices with $U/t=8$ and $n_h=1/8$ hole doping. The right table shows results for the $8 \times 8$ torus with ViT and Symm-ViT-Ti architectures under the same physical parameters. Energies are estimated using $131072$ samples.}
\label{tab:combined_energy}
\end{table}

\section{Additional results for the $8 \times 8$}
\begin{figure}[th!]
    \centering
    \includegraphics[width=\linewidth]{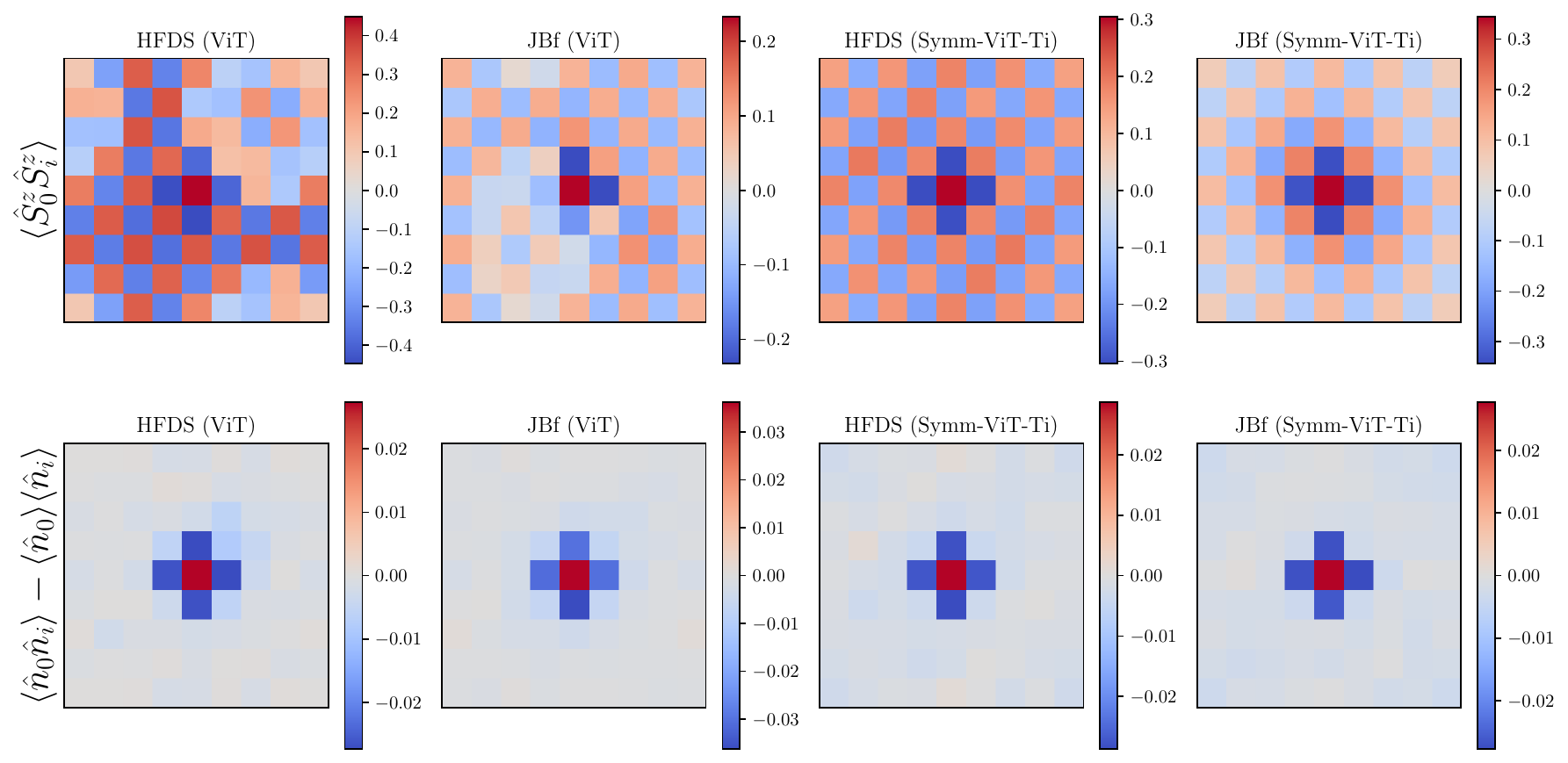}
    \caption{Results for the physical observables on the $8 \times 8$ lattice for $U/t=8$ and $n_h=1/8$. The top row shows the spin-spin correlation in real space for the HFDS-ViT and JBf wavefunctions parameterized by an unsymmetrized ViT and the Symm-ViT-Ti. The bottom row shows the density-density correlation function.}
    \label{fig:placeholder}
\end{figure}

\section{Test patching strategies}
\label{appedix: Patching}
We conducted a series of experiments to investigate how different patching strategies influence model performance. In particular, we compared a 2×2 patching scheme where local groups of orbitals are treated as input units to a baseline where each orbital is treated as an individual patch (1×1). This allows us to assess whether incorporating short-range spatial structure improves the model’s ability to capture relevant correlations. 
We also examined the effect of how spin information is represented, by testing whether concatenating the spin up and spin down subsectors as separate channels of a single input offers advantages over treating them as independent inputs. These experiments provide insight into how local context and spin representation affect the expressiveness and efficiency of the network.

\begin{figure*}[th!]
\centering    \includegraphics[width=\linewidth]{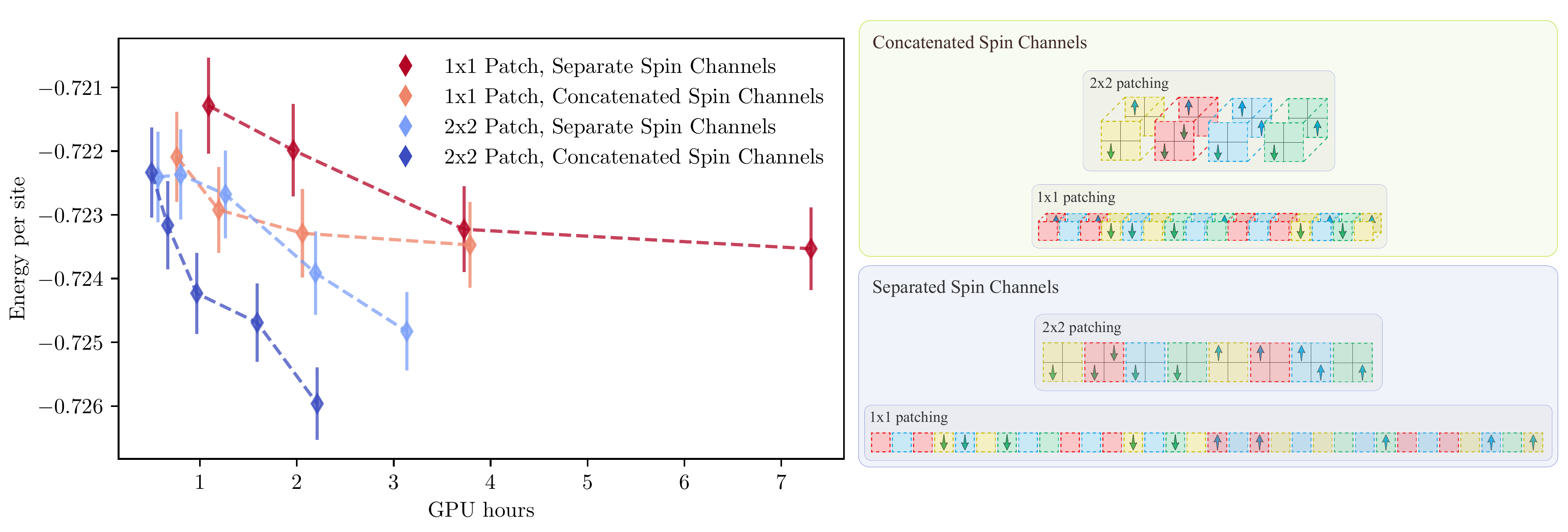}
    \caption{Left: Effect of different patching strategies on a $4\times 8$ PBC–OBC system at $U=8$, with each point corresponding to a different number of encoder blocks. Right: Schematic of the patching schemes. Top row: 2×2 and 1×1 patches with spin concatenated; bottom row: same patches with spin treated separately. The best trade-off between energy and GPU time uses 2×2 patches with concatenated spin.}
    \label{fig:patching_scheme}
\end{figure*}

\section{Role of the Jastrow Factor in the HFDS Ansatz}
\label{appedix:jastrow_hfds}

In the main text, we compare the performance of the JBf with the HFDS ansätze. While the JBf ansatz explicitly includes a density-density Jastrow factor to capture electronic correlations, the HFDS ansatz does not. This appendix clarifies this architectural choice, demonstrating both theoretically and empirically that an explicit Jastrow factor is redundant for the HFDS.

From a theoretical perspective, the HFDS ansatz inherently subsumes Jastrow-like correlations. As demonstrated in Ref.~\cite{robledo2022fermionic}, the HFDS is a strict generalization of the Slater-Jastrow wavefunction. In the limit of a single hidden fermion ($\tilde{N} = 1$), it exactly recovers the Slater-Jastrow structure. Since our implementation utilizes $\tilde{N} \ge 1$ and unrestricted matrices, the ansatz naturally spans this variational space without requiring manual augmentation.

\begin{figure}[h]
    \centering
    \includegraphics[width=\columnwidth]{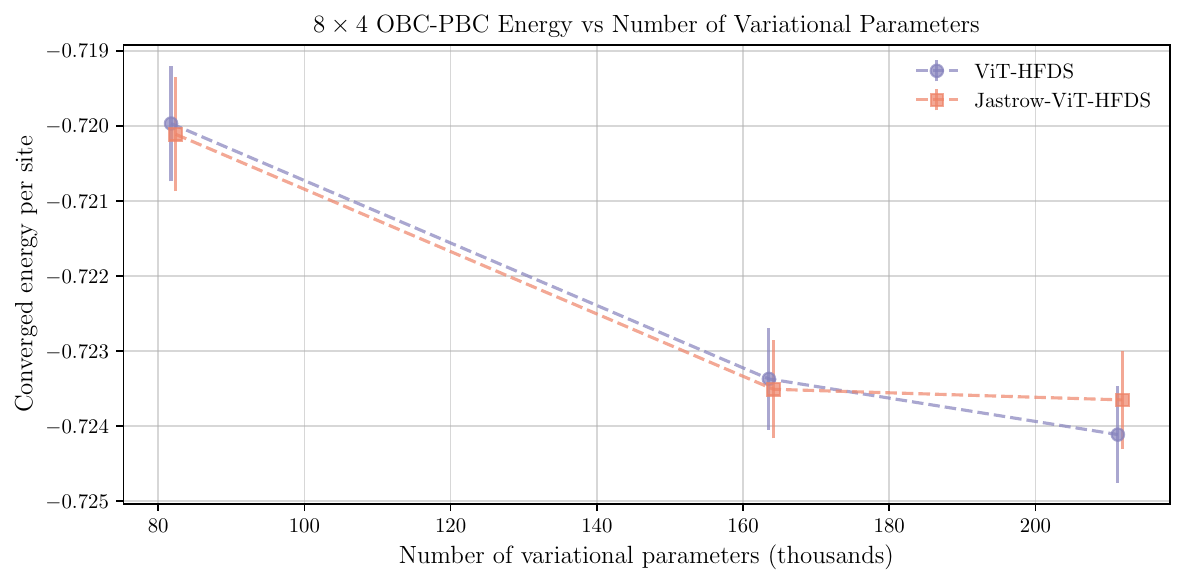}
    \caption{Comparison of the variational energy achieved by the HFDS ansatz with (coral) and without (purple) an explicit two-body exponential Jastrow factor. The inclusion of the explicit factor provides no significant advantage.}
    \label{fig:hfds_jastrow_compare}
\end{figure}

Furthermore, as established in Ref.~\cite{liu2024unifying}, the HFDS architecture inherently includes a multiplicative symmetric scalar term:
\begin{equation}
    \Psi_{r-HF}(c) = \det(F_\theta(c)) \cdot \det\left( \left( A - B F_\theta(c)^{-1} C_\theta(c) \right)[c] \right).
\end{equation}
The first term, $\det(F_\theta(c))$, depends on the physical coordinates $c$ exclusively through permutation-invariant operations, guaranteeing that the overall anti-symmetry is preserved. Thus, it acts as a symmetric scalar pre-factor that effectively fulfills the exact role of a Jastrow factor.\\
To provide direct empirical confirmation, we performed additional benchmarks comparing the baseline HFDS ansatz against an HFDS augmented with an explicit two-body exponential Jastrow factor. As shown in Fig.~\ref{fig:hfds_jastrow_compare}, supplementing the HFDS wavefunction with this explicit factor offers no significant improvement in the variational energy. This indicates that the neural network capacity is already sufficient to learn the necessary density-density correlations directly through the hidden fermion determinant.

\end{document}